\documentclass[10pt,conference]{IEEEtran}

\makeatletter
\def\endthebibliography{%
	\def\@noitemerr{\@latex@warning{Empty `thebibliography' environment}}%
	\endlist
}
\makeatother

\usepackage{graphicx,epsfig,amssymb,amsmath,url,latexsym,stfloats,array,bm,bbm}
\usepackage[tight,footnotesize]{subfigure}
\usepackage{makecell}
\usepackage{mathrsfs}
\usepackage{amsthm}
\usepackage{longtable}
\usepackage{amsfonts}
\usepackage{caption}
\usepackage{graphicx}
\usepackage{multirow}
\usepackage{longtable}
\usepackage{afterpage}
\usepackage{subeqnarray}
\usepackage{epstopdf}
\usepackage{empheq}
\usepackage{latexsym}
\usepackage{color, soul}
\usepackage{framed}
\usepackage{lipsum}
\usepackage{color}
\usepackage{stfloats}
\usepackage{setspace}
\definecolor{shadecolor}{rgb}{1,0,0}
\usepackage{cite}
\usepackage{hyperref}
\usepackage{verbatim}
\usepackage{algorithmic}
\usepackage[linesnumbered,ruled]{algorithm2e}
\usepackage[hang]{footmisc}
\usepackage{caption}
\usepackage{cuted}
\begin{document}
\title{
Digital Twin-Assisted Collaborative Transcoding for Better User Satisfaction in Live Streaming}
{\setstretch{1.0}
	\author{
	\IEEEauthorblockN{Xinyu Huang\IEEEauthorrefmark{1}, Mushu Li\IEEEauthorrefmark{2},  Wen Wu\IEEEauthorrefmark{3}, Conghao Zhou\IEEEauthorrefmark{1}, and Xuemin (Sherman) Shen\IEEEauthorrefmark{1}}
	    \IEEEauthorblockA{\IEEEauthorrefmark{1}Department~of~Electrical~\&~Computer~Engineering,~University~of~Waterloo,~Canada
	    \\\IEEEauthorrefmark{2}Department of Electrical, Computer, and Biomedical Engineering, Toronto Metropolitan University, Canada
	    \\\IEEEauthorrefmark{3}Frontier Research Center, Peng~Cheng~Laboratory,~China
	    \\Email: \{x357huan, c89zhou, sshen\}@uwaterloo.ca, mushu1.li@ryerson.ca, wuw02@pcl.ac.cn}
			}
}

\maketitle
\pagestyle{empty}  
\thispagestyle{empty} 
\begin{abstract}
In this paper, we propose a digital twin (DT)-assisted cloud-edge collaborative transcoding scheme to enhance user satisfaction in live streaming. We first present a DT-assisted transcoding workload estimation (TWE) model for the cloud-edge collaborative transcoding. Particularly, two DTs are constructed for emulating the cloud-edge collaborative transcoding process by analyzing spatial-temporal information of individual videos and transcoding configurations of transcoding queues, respectively. Two light-weight Bayesian neural networks are adopted to fit the TWE models in DTs, respectively. We then formulate a transcoding-path selection problem to maximize long-term user satisfaction within an average service delay threshold, taking into account the dynamics of video arrivals and video requests. The problem is transformed into a standard Markov decision process by using the Lyapunov optimization and solved by a deep reinforcement learning algorithm. Simulation results based on the real-world dataset demonstrate that the proposed scheme can effectively enhance user satisfaction compared with benchmark schemes.

\end{abstract}

\section{Introduction}
With the prevalence of smart mobile equipment and ubiquitous Internet access, an increasing number of amateur or professional broadcasters start to utilize online video platforms, such as Youtube Live, Facebook Live, Twitch TV, etc., to produce live streams and interact with users anytime and anywhere. According to a recent Grand View Research report, the live streaming market is expected to increase from \$70 billion in 2021 to about \$224 billion in 2028 \cite{report}. As a critical technology to guarantee continuous and high-definition video playback, video transcoding encodes video streams pipelined at online video platforms into multiple video versions in real time \cite{Transcode}. Since a user's device has a small buffer size for live streaming playback, the performance of video transcoding can directly affect user satisfaction. Video transcoding can be conducted in a cloud server, i.e., cloud-transcoding, and an edge server, i.e., edge-transcoding \cite{shi1}. However, video transcoding is a computation-intensive process, and the transcoding delay dominates the service delay, almost $70 \%$ \cite{delay}. Relying solely on cloud-transcoding or edge-transcoding can incur a large transmission delay and heavy computation overhead, respectively \cite{When,shi2}. Therefore, efficient cloud-edge collaborative transcoding has attracted considerable research attention. 

In the literature, significant efforts have been devoted to improve the cloud-edge collaborative transcoding performance. Pang \textit{et al}. proposed to dispatch newly generated video streams to appropriate transcoding queues (TQs) in cloud and edge servers based on predicted interaction intensities in different streaming channels, which can satisfy users’ heterogeneous quality of experience (QoE) requirements \cite{MM}. To reduce collaborative transcoding cost, Erfanian \textit{et al}. constructed a transcoding-cost-based multicast tree for each cloud and edge computing server to determine where and how many TQs should be deployed \cite{OSCAR}. Zhu \textit{et al}. proposed an auction-based approach for TQ selection to further reduce transcoding cost within a prescribed transcoding delay threshold \cite{When}. The above works utilize a general transcoding workload estimation (TWE) model for different kinds of video streams in different TQs. However, the TWE model ignores videos' spatial-temporal information and servers' transcoding configurations, thus bringing TQs' length estimation errors. Such estimation errors can result in improper transcoding-path selections for video streams, thereby rendering user satisfaction degradation. Hence, it is paramount to construct an accurate TWE model. 

The digital twin (DT) technology is a potential solution since it is a digital representation of a physical entity (PE) that can accurately reflect its status and feature via real-time synchronization between the DT and PE \cite{Holistic}. The DT technology can be utilized to store and analyze users' data to construct user-specific and real-time personalized QoE models for tailored network resource management \cite{PQoE}. We leverage the DT technology to construct virtual TQs based on estimated transcoding workloads and transcoding-path selections, which can reflect the dynamics of physical TQs and emulate transcoding performance of physical TQs in real time.

In this paper, we present a DT-assisted cloud-edge collaborative transcoding scheme for live streaming services to enhance long-term user satisfaction given an average service delay threshold. Specifically, we first propose a DT-assisted collaborative transcoding model to capture the dynamics of physical TQs in cloud and edge servers. The cloud and edge transcoding DTs are established at the network controller, consisting of historical transcoding data, a TWE model and virtual TQs. The TWE model adopts a Bayesian neural network (BNN) to fit historical transcoding data in order to reduce estimation errors. Based on the TWE model, the virtual TQs are constructed to reflect the dynamics of physical TQs. Secondly, we design a tailored deep reinforcement learning (DRL) algorithm for making transcoding-path selections. The objective is to maximize long-term user satisfaction within an average service delay threshold by optimizing transcoding-path selections. Since the formulated transcoding-path selection problem is a constrained Markov decision process (CMDP), we leverage the Lyapunov optimization technique to transform it into a standard MDP. Extensive experiments conducted on the real-world dataset demonstrate that the proposed DT-assisted collaborative transcoding scheme can effectively enhance user satisfaction within an service delay threshold compared with benchmarks. The main contributions of this paper are summarized as follows:
\begin{itemize}
	\item[$\bullet$] We propose a DT-assisted collaborative transcoding model, which can reduce the TQ length estimation error and capture its dynamics.
	
	\item[$\bullet$] We develop a cloud-edge collaborative transcoding algorithm based on DRL, which can realize an online transcoding-path selection.
	
\end{itemize}

The remainder of this paper is organized as follows. The DT-assisted collaborative transcoding scheme is presented in Section \ref{system}. The DRL-based transcoding scheduling algorithm is proposed in Section \ref{Solution}. Simulation results are provided in Section \ref{Result}, followed by the conclusion in Section \ref{Conclusion}.

\section{DT-Assisted Collaborative Transcoding Scheme}\label{system}
\begin{figure}[t]
	\centering
	\includegraphics[width=8.8cm]{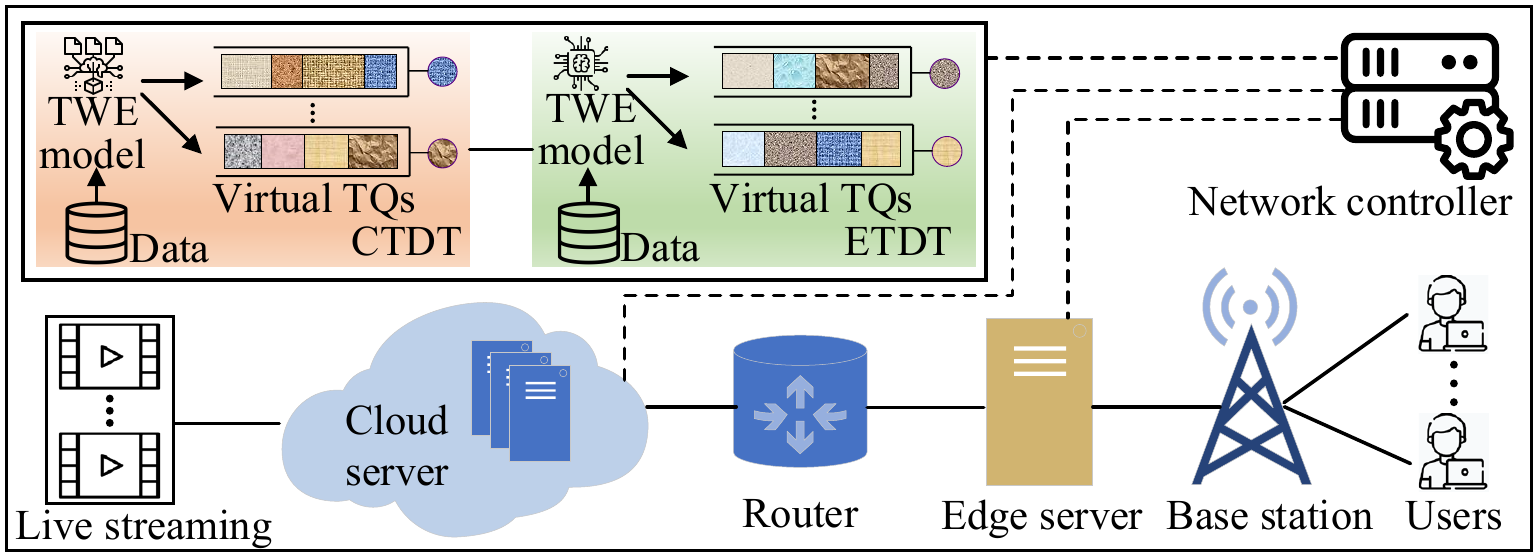}
	\caption{DT-assisted collaborative transcoding scenario.}
	\label{fig:video_services}
\end{figure}
\subsection{System Model}
As shown in Fig.~\ref{fig:video_services}, we consider a DT-assisted collaborative transcoding scenario, which mainly consists of a cloud server, an edge server, and DTs. 

Users' devices generate video requests to the network controller for requesting live streams to be played soon with different video qualities every 0.5 s. The cloud and edge servers collaboratively transcode requested live streams and deliver them to users' devices. Specifically, the cloud server is deployed with three kinds of TQs\footnote{Our proposed scheme can also be applied to the scenario where the number of TQs in the cloud or edge server is more than three.} for video transcoding based on different encoding presets\footnote{https://trac.ffmpeg.org/wiki/Encode/H.264\\ https://trac.ffmpeg.org/wiki/Encode/H.265}, i.e., slow TQ, medium TQ, and fast TQ. From the fast to slow TQs, the transcoding speed gradually decreases but the transcoded video quality gradually rises. A higher video quality usually corresponds to a larger video bit rate. The edge server is deployed with two kinds of TQs, i.e., medium TQ and fast TQ, considering that the edge server with limited computing resources hardly supports a slow TQ that requires plenty of computing resources. As the minimal transcoding unit of video streams, a group of pictures (GoP) is dispatched to different TQs in the cloud and edge servers to generate multiple video versions of different qualities. When a GoP is delivered from the cloud server to a user's device, there exist two kinds of transcoding paths, i.e., one-step transcoding path and two-step transcoding path, according to the queues that the GoP goes through. The former path refers to the original GoP being transcoded in any cloud TQ and then directly sent to users. The latter path indicates that the original GoP is first transcoded in the high or medium cloud TQ, and then dispatched to any edge TQ for further transcoding. By transcoding the GoP into multiple video versions, users’ differentiated requests can be satisfied.

To properly select the video transcoding path, two kinds of DTs are constructed, i.e., cloud transcoding DT (CTDT) and edge transcoding DT (ETDT), and located at the network controller. Both CTDT and ETDT consist of three components: historical transcoding data on GoPs' spatial-temporal information and servers' transcoding configurations, a unique TWE model and multiple virtual TQs. The TWE model and virtual TQs are used to characterize the transcoding workload of each GoP and the dynamics of physical TQs, respectively. Specifically, the TWE model adopts a BNN to analyze the historical transcoding data, which can estimate the specific transcoding workload for each GoP in different TQs. To ensure the estimation accuracy of TWE models, actual transcoding data are recorded in the cloud and edge servers if the transcoding workload bias exceeds a prescribed threshold. The recorded transcoding data are periodically uploaded to the network controller for the BNN model update. In addition, the virtual TQs in both CTDT and ETDT are digital representations of physical TQs in the cloud and edge servers, which are updated based on selected transcoding paths and estimated transcoding workloads of GoPs. The cloud and edge servers also send the queue synchronization message if the TQ length bias exceeds a prescribed threshold. 

When users send GoP requests consisting of GoP indexes and corresponding video bit rates to the network controller, CTDT and ETDT collect GoPs' transcoding data, and use respective TWE models to estimate corresponding transcoding workloads in different TQs. Then, the network controller makes the transcoding-path selection for each GoP by taking GoP requests, estimated transcoding workloads, and virtual TQ lengths into account. TWE models and virtual TQs are updated based on the recorded transcoding data and determined transcoding paths of GoPs, respectively.

\subsection{DT-Assisted GoP's TWE}
CTDT and ETDT can emulate the dynamics of physical TQs for newly arrived GoPs by analyzing the GoP's spatial-temporal information, the servers' transcoding configurations, and the computing capability of each TQ. 

\textbf{Spatial and temporal information:} The spatial information (SI) and temporal information (TI) are critical factors that can affect the transcoding workload \cite{STI}. The SI indicates the amount of spatial details in a video frame. Specifically, each video frame in a GoP is first filtered by the Sobel filter, and then the standard deviation over the pixels in each filtered frame is calculated. The maximum standard deviation among filtered frames represents the GoP's SI. Therefore, the SI of GoP $k$ can be calculated by
\begin{equation}\label{SI}
	{{\varsigma }_{k}}=\underset{l\in \mathcal{L}}{\mathop{\max }}\,\left\{ \sigma \left[ \Theta (F_{k}^{l}) \right] \right\},
\end{equation}
where $F_{k}^{l}$ is video frame $l$ of GoP $k$, and $\Theta (\cdot )$ is the Sobel filter operation. In addition, $\sigma $ is the standard deviation of a filtered frame, and $\mathcal{L}$ is the set of GoP frames.  

The TI indicates the amount of temporal changes of a video frame sequence. Specifically, the pixel difference of each adjacent frames is first calculated. Then, the standard deviation of each pixel difference is calculated. The maximum standard deviation is chosen to represent the GoP's TI. Therefore, the TI of GoP $k$ can be calculated by
\begin{equation}\label{TI}
	{{\xi }_{k}}=\underset{l\in \mathcal{L}\setminus\{L\}}{\mathop{\max }}\,\left\{ \sigma \left[ F_{k}^{l+1}-F_{k}^{l} \right] \right\}.
\end{equation}

\textbf{TWE:} To emulate the transcoding performance of each TQ for newly arrived GoPs, we select the BNN to construct TWE models in both CTDT and ETDT. The BNN is a widely used tool to fit the function by finding the distribution of the weighting vector and uses the regularization technology to avoid overfitting \cite{BNN}.

The input of the BNN-based model is a multi-dimension vector, denoted by ${{V}}$, which mainly abstracts a GoP's spatial-temporal information, servers' transcoding configurations, and TQs' computing capabilities, as presented in Table~\ref{Table1}. 
\begin{table}[t]
	\footnotesize
	\centering
	\captionsetup{justification=centering, singlelinecheck=false}
	\caption{The input vector of the BNN-based DT model}
	\label{Table1}
	\begin{tabular}{|c|c|c|c|}
		\hline\hline
		\textbf{Index} & \textbf{Name}     & \textbf{Index} & \textbf{Name}       \\ \hline
		1              & Encoding Preset & 5              & Computing Processor \\ \hline
		2              & Number of Frames  & 6              & Computing Density   \\ \hline
		3              & Resolution        & 7              & Computing Capability  \\ \hline
		4              & SI                & 8              & TI                  \\ \hline
	\end{tabular}
\end{table}

The output of the BNN-based model is the estimated transcoding workload for a GoP in TQ $i$, which can be expressed as ${{\Omega }_{i}}\left( {{V}} \right)$. The function ${{\Omega }_{i}}(\cdot )$ is fitted based on the BNN regularization algorithm and updated periodically to decrease the estimation error. The recorded actual transcoding workloads are selected as ground truths to calculate the mean squared error (MSE) for model training and update.

\subsection{DT-Assisted Collaborative Transcoding Model}
\begin{figure}[t]
	\centering
	\includegraphics[width=9cm]{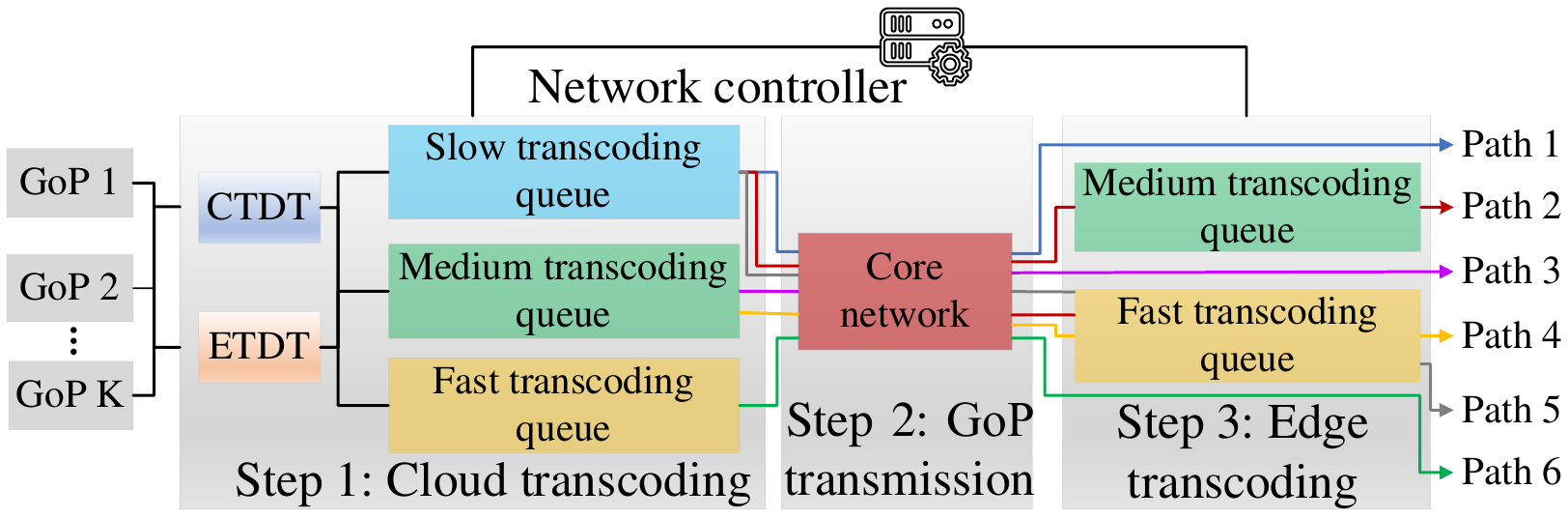}
	\caption{Collaborative transcoding procedure.}
	\label{fig:collaborative}
\end{figure}

After estimating transcoding workloads of GoPs in different TQs in the cloud server and edge server, the network controller will select the appropriate transcoding path for each GoP based on GoP requests, estimated transcoding workloads, and virtual TQ lengths. A simplified procedure is shown in Fig.~\ref{fig:collaborative}. There exist six transcoding paths, including three one-step transcoding paths, i.e., path 1: slow cloud-transcoding, path 3: medium cloud-transcoding, and path 6: fast cloud-transcoding, and three two-step transcoding paths, i.e., path 2: slow-medium cloud-edge transcoding, path 4: medium-fast cloud-edge transcoding, and path 5: slow-fast cloud-edge transcoding.

The collaborative transcoding decision is made in each scheduling slot, indexed by $t$. We consider that there are $K$ GoPs arriving at the cloud server at scheduling slot $t$, and the corresponding set of GoPs is $\mathcal{K}$. The sets of cloud TQs and edge TQs are denoted by ${{\Lambda }_{1}}=\left\{ 1,2,3 \right\}$ and ${{\Lambda }_{2}}=\left\{ 4,5 \right\}$, respectively, which correspond to the sets of CTDT queues and ETDT queues. To determine the transcoding path for each GoP, we define a binary transcoding position variable $x_{t,k}^{i}$, where $x_{t,k}^{i}=1$ indicates that GoP $k$ is dispatched to TQ $i$ for transcoding at scheduling slot $t$; Otherwise, $x_{t,k}^{i}=0$. The set of index $i$ ranges from $1$ to $5$, which refers to the slow, medium, and fast CTDT queues, and the medium and fast ETDT queues, respectively. 

\textbf{CTDT queue dynamics:}
In scheduling slot $t$, the queue length of TQ $i$ in CTDT, denoted by, $L_{t}^{i}$, is updated via
\begin{equation}
	L_{t+1}^{i}={{\left[ L_{t}^{i}+\sum\limits_{k=1}^{K}{x_{t,k}^{i}\frac{{{\Omega }_{1}}\left( V_{t}^{k} \right)b_{t}^{k}}{{{f}_{i}}{{\kappa }_{i}}}}-d \right]}^{+}},\forall i\in {{\Lambda }_{1}},
\end{equation}
where ${\Omega }_{1}\left( V_{t}^{k} \right)$ is the estimated transcoding workload for GoP $k$ in TQ 1, and $b_{t}^{k}$ is the original bit rate of GoP $k$ at scheduling slot $t$. Parameters ${{f}_{i}}$ and ${{\kappa }_{i}}$ represent the computing capability of TQ $i$ and the corresponding computing density, respectively. Parameter $d$ is the scheduling slot length. 

\textbf{ETDT queue dynamics:} To reflect the queue dynamics of ETDT queues, we count how many GoPs are dequeued in each CTDT queue and sent to each ETDT queue based on two-step transcoding paths in each scheduling slot. For the medium ETDT queue, i.e., TQ 4, enqueued GoPs origin from path 2, and the corresponding queue length is updated via
\begin{equation}
	L_{t+1}^{4}={{\left[ L_{t}^{4}+\frac{\alpha _{t}^{1}\min \left\{ L_{t}^{1},T \right\}\overline{\Omega }_{t}^{1\to 4}{{f}_{1}}{{\kappa }_{1}}}{\overline{\Omega }_{t}^{1}{{f}_{4}}{{\kappa }_{4}}}-d \right]}^{+}},
\end{equation}
where $\overline{\Omega }_{t}^{1}$ is the average transcoding workload of GoPs in the slow CTDT queue at scheduling slot $t$. Here, $\alpha _{t}^{1}$ indicates the ratio of GoPs of path 2 in the slow CTDT queue, i.e., TQ 1, which is updated via
\begin{equation}
	\alpha _{t+1}^{1}=\frac{\alpha _{t}^{1}L_{t}^{1}+\sum\nolimits_{k=1}^{K}{x_{t,k}^{4}\frac{{{\Omega }_{1}}\left( V_{t}^{k} \right)b_{t}^{k}}{{{f}_{1}}{{\kappa }_{1}}}}}{L_{t}^{1}+\sum\nolimits_{k=1}^{K}{x_{t,k}^{1}\frac{{{\Omega }_{1}}\left( V_{t}^{k} \right)b_{t}^{k}}{{{f}_{1}}{{\kappa }_{1}}}}}.
\end{equation}
In addition, $\overline{\Omega }_{t}^{1\to 4}$ is the average transcoding workload of GoPs of path 2 in TQ 1, which is updated via
\begin{equation}
\overline{\Omega }_{t+1}^{1\to 4}=\frac{\sum\nolimits_{k=1}^{K}{x_{t,k}^{1}x_{t,k}^{4}{{\Omega }_{4}}(V_{t}^{k})b_{t}^{k}}+\overline{\Omega }_{t}^{1\to 4}B_{t}^{1\to 4}}{\sum\nolimits_{k=1}^{K}{x_{t,k}^{1}x_{t,k}^{4}s_{t}^{k}}+B_{t}^{1\to 4}}, 	
\end{equation}
where $B_{t}^{1\to 4}$ is the bit rate of all GoPs belonging to path 2 at scheduling slot $t$.

For the fast ETDT queue, the enqueued GoPs can be sent from path 4 and path 5. Therefore, the fast ETDT queue, i.e., TQ 5, is updated via	
\begin{equation}
	\begin{split}
	L_{t+1}^{5}=&{{\left[ L_{t}^{5}+\frac{\alpha _{t}^{2}\min \left\{ L_{t}^{1},T \right\}\overline{\Omega }_{t}^{1\to 5}{{f}_{1}}{{\kappa }_{1}}}{\overline{\Omega }_{t}^{1}{{f}_{5}}{{\kappa }_{5}}}\right.}}	\\&	
	{{\left.+\frac{\alpha _{t}^{3}\min \left\{ L_{t}^{2},T \right\}\overline{\Omega }_{t}^{2\to 5}{{f}_{2}}{{\kappa }_{2}}}{\overline{\Omega }_{t}^{2}{{f}_{5}}{{\kappa }_{5}}}-d \right]}^{+}},
	\end{split}
\end{equation}
where $\overline{\Omega }_{t}^{2}$ is the average transcoding workload of GoPs in the medium CTDT queue, i.e., TQ 2, at scheduling slot $t$. Here, $\alpha _{t}^{2}$ and $\alpha _{t}^{3}$ refer to the ratio of GoPs of path 4 and 5 in TQ 1 and TQ 2, respectively, which are updated via
\begin{equation}
	\alpha _{t+1}^{i}=\frac{\alpha _{t}^{i}L_{t}^{i}+\sum\nolimits_{k=1}^{K}{x_{t,k}^{5}\frac{{{\Omega }_{i-1}}(V_{t}^{k}){{b}_{t,k}}}{{{f}_{i-1}}{{\kappa }_{i-1}}}}}{L_{t}^{i-1}+\sum\nolimits_{k=1}^{K}{x_{t,k}^{i-1}\frac{{{\Omega }_{i-1}}(V_{t}^{k}){{b}_{t,k}}}{{{f}_{i-1}}{{\kappa }_{i-1}}}}},\forall i\in \left\{ 2,3 \right\}.
\end{equation}
In addition, $\overline{\Omega }_{t}^{1\to 5}$ and $\overline{\Omega }_{t}^{2\to 5}$ indicate the average transcoding workloads of GoPs of path 4 and path 5 in TQ 1 and TQ 2, respectively, which are updated by 
\begin{equation}
	\overline{\Omega }_{t+1}^{i\to 5}\!=\!\frac{\sum\nolimits_{k=1}^{K}{x_{t,k}^{i}\!x_{t,k}^{5}{{\Omega }_{5}}\left( V_{t}^{k} \right){{b}_{t,k}}}+\overline{\Omega }_{t}^{i\to 5}B_{t}^{i\to 5}}{\sum\nolimits_{k=1}^{K}{x_{t,k}^{i}x_{t,k}^{5}{{b}_{t,k}}}+B_{t}^{i\to 5}},\forall i\!\in\! \left\{ 1,2 \right\}.
\end{equation}

\textbf{Service delay:} Based on the analysis of queuing dynamics in CTDT and ETDT, we can estimate the service delay, $D_t(X_t)$, in this cloud-edge collaborative transcoding system, which is given by
\begin{equation}
	{{D}_{t}}({{X}_{t}})=\frac{1}{3}\sum\nolimits_{i\in {{\Lambda }_{1}}}{L_{t}^{i}}+{{I}_{t}}+\frac{1}{2}\sum\nolimits_{i\in {{\Lambda }_{2}}}{L_{t}^{i}},
\end{equation}
where ${{X}_{t}}$ is the decision variable set, and ${{X}_{t}}={{\left\{ x_{t,k}^{i} \right\}}_{i\in \mathcal{I},k\in \mathcal{K}}}$. Here, ${{I}_{t}}$ is the transmission delay between the cloud server and the edge server at scheduling slot $t$, which can be estimated based on the round-trip time (RTT).

\textbf{User satisfaction:} In addition to the service delay, we also introduce user satisfaction to evaluate transcoding performance. The user satisfaction, $W_t$, refers to the ratio of users' video requests that can be satisfied through transcoding operations in scheduling slot $t$, which is depicted as
\begin{equation}
	{{W}_{t}}=\frac{\sum\nolimits_{k=1}^{K}{
				x_{t,k}^{1}w_{t,k}^{1}+\left( x_{t,k}^{2}+x_{t,k}^{1}x_{t,k}^{4} \right)w_{t,k}^{2}+ M_{t,k}}}
	{\sum\nolimits_{k=1}^{K}{w_{t,k}^{1}+w_{t,k}^{2}+w_{t,k}^{3}}},
\end{equation}
where $w_{t,k}^{1},w_{t,k}^{2},w_{t,k}^{3}$ represent the number of GoP requests of high, medium, and low quality, respectively. Here, $M_{t,k} = \left( x_{t,k}^{3}+x_{t,k}^{1}x_{t,k}^{5}+x_{t,k}^{2}x_{t,k}^{5} \right)w_{t,k}^{3}$.

\subsection{Problem Formulation}
Our objective is to maximize long-term user satisfaction within an average service delay threshold over $T$ scheduling slots. Correspondingly, the optimization problem is formulated as
\begin{align}\label{sat}
	{{\textbf{P}}_{\text{0}}:} \quad & \underset{{{X}_{t}}}{\mathop{\max }}\,\underset{T\to \infty }{\mathop{\lim }}\,\frac{1}{T}\sum\limits_{t=1}^{T}{{{W}_{t}}({{X}_{t}})}, \\ 
	\text{s}\text{.t}\text{. }\quad&\underset{T\to \infty }{\mathop{\lim }}\,\frac{1}{T}\sum\limits_{t=1}^{T}{{{D}_{t}}({{X}_{t}})}\le \overline{D}, \tag{12a} \label{1a} \\ 
	& x_{t,k}^{2}+x_{t,k}^{4}\le 1,\tag{12b} \label{1b} \\ 
	& x_{t,k}^{3}+x_{t,k}^{5}\le 1,\tag{12c}\label{1c} \\ 
	& x_{t,k}^{1}-x_{t,k}^{4}\ge 0,\tag{12d} \label{1d}\\ 
	& x_{t,k}^{1}+x_{t,k}^{2}-x_{t,k}^{5}\ge 0,\tag{12e} \label{1e}\\ 
	& x_{t,k}^{i}\in \left\{ 0,1 \right\},\forall i\in \mathcal{I}\text{,} \tag{12f}
\end{align}
where $\overline{D}$ is the average service delay threshold. Constraint~(\ref{1a}) represents the average service delay requirement. Constraints~(\ref{1b},\ref{1c}) avoid the repeated transcoding operation for a GoP between cloud TQs and edge TQs. Constraint~(\ref{1d}) guarantees that a GoP is transcoded from the high-quality version to the medium-quality version. Constraint~(\ref{1e}) guarantees that a GoP is transcoded from the high-quality and medium-quality versions to the low-quality version. 

The formulated problem is a CMDP since the state transition is Markovian with the long-term constraint. However, the CMDP cannot be directly solved by general DRL algorithms due to the intractable long-term constraint \cite{wu}. Therefore, we first transform it into a standard MDP and then develop a DRL algorithm to solve it.

\section{DRL-Based Transcoding Scheduling Algorithm}\label{Solution}
\subsection{Problem Transformation}
The Lyapunov optimization technique is an effective way to handle a long-term constraint \cite{Lya} in an optimization problem. The main idea is to establish a deficit queue of service delay to characterize the satisfaction status of the long-term constraint, which can transform the time average constraint problem into a queue stability problem.

First, we establish the deficit queue of service delay, as follows:
\begin{equation}
	{{Z}_{t+1}}={{\left[ {{D}_{t}}\left( {{X}_{t}} \right)-\overline{D}+{{Z}_{t}} \right]}^{+}},
\end{equation}
where ${{Z}_{t}}$ represents the deviation of achieved instantaneous values from the long-term constraint, whose initial state is set to 0. To characterize the satisfaction status of the long-term constraint, the Lyapunov function is defined as $L({{Z}_{t}})=\frac{1}{2}{{({{Z}_{t}})}^{2}}$\cite{Lya}. A smaller Lyapunov function value indicates a better satisfaction of the long-term constraint.

Second, to guarantee that the Lyapunov function can be consistently preserved within a small value, the one-shot Lyapunov drift is introduced to capture the variation of Lyapunov function values across two subsequent time slots. Therefore, the one-shot Lyapunov drift is defined as $\Delta ({{Z}_{t}})=L({{Z}_{t+1}})-L({{Z}_{t}})$. The upper bound of $\Delta ({{Z}_{t}})$ can be derived as
\begin{equation}\small
\begin{split}
	& \Delta ({{Z}_{t}})=\frac{1}{2}{{({{Z}_{t+1}})}^{2}}-\frac{1}{2}{{({{Z}_{t}})}^{2}} \\ 
	& \le \frac{1}{2}{{\left( {{Z}_{t}}+{{D}_{t}}\left( {{X}_{t}} \right)-\overline{D} \right)}^{2}}-\frac{1}{2}{{\left( {{Z}_{t}} \right)}^{2}} \\
	 & ={{Z}_{t}}\left( {{D}_{t}}\left( {{X}_{t}} \right)-\overline{D} \right)+\frac{1}{2}{{\left( {{D}_{t}}\left( {{X}_{t}} \right)-\overline{D} \right)}^{2}} \\  
	& ={{Z}_{t}}\left( {{D}_{t}}\left( {{X}_{t}} \right)-\overline{D} \right)+\frac{1}{2}{{\left( \frac{1}{3}\sum\limits_{i=1}^{3}{L_{t}^{i}}+{{I}_{t}}+\frac{1}{2}\sum\limits_{i=4}^{5}{L_{t}^{i}}-\overline{D} \right)}^{2}} \\ 
	& \le {{Z}_{t}}\left( {{D}_{t}}\left( {{X}_{t}} \right)-\overline{D} \right)+\Gamma,
\end{split}
\end{equation}
where $\Gamma = \frac{1}{2}{{\left( \frac{1}{3}\sum\limits_{i=1}^{3}{L_{i}^{\max }}+{{I}_{\max }}+\frac{1}{2}\sum\limits_{i=4}^{5}{L_{i}^{\max }}-\overline{D} \right)}^{2}}$ and it is a constant. Here, $L_{i}^{\max }$ and ${{{I}}_{\max }}$ are the maximum length of TQ $i$ and the maximum RTT, respectively. 

Third, the original optimization problem of maximizing long-term user satisfaction within an average service delay threshold can be transformed to minimize a drift-plus-cost. Problem $\textbf{P}_0$ is reformulated as
\begin{equation}\label{trans}
	\begin{split}
 \textbf{P}_{\text{1}}:&\quad\underset{{{X}_{t}}}{\mathop{\min }}\,{{Z}_{t}}\left( D\left( {{X}_{t}} \right)-\overline{D} \right)-V{{W}_{t}}({{X}_{t}}), \\ 
	\text{s}\text{.t}\text{. }&(\ref{1b}),(\ref{1c}),(\ref{1d}),(\ref{1e}), \\ 
	& x_{t,k}^{i}\in \left\{ 0,1 \right\},\forall i\in \mathcal{I}.
	\end{split}
\end{equation}

\subsection{Proposed Algorithm}
The transformed problem is a nonlinear integer programming problem, which is hard to be directly solved. Since the state transition is Markovian, we adopt the dueling DQN (DDQN) algorithm \cite{DDQN} to obtain the transcoding-path selection for each GoP in each scheduling slot. Specifically, the dueling architecture constructs two streams of fully connected layers to provide separate estimations of value and advantage functions, which can learn which TQ lengths are valuable without having to learn the effect of transcoding-path selection of each GoP.

The state includes TQs' lengths, the deficit queue length, estimated transcoding workloads, and bit rates of GoPs, i.e.,
${{s}_{t}}=\left\{ {{\left\{ L_{t}^{i} \right\}}_{i\in \mathcal{I}}},Z_t, {{\left\{ {{\Omega }_{i}}\left( V_{t}^{k} \right) \right\}}_{i\in \mathcal{I},k\in \mathcal{K}}},{{\left\{ {{b}_{t,k}} \right\}}_{k\in \mathcal{K}}} \right\}$. The action includes all transcoding decisions in Eq.~(\ref{trans}), i.e., ${{a}_{t}}={{\left\{ x_{t,k}^{i} \right\}}_{_{i\in \mathcal{I},k\in \mathcal{K}}}}$. The reward at step $t$ is the opposite value of drift-plus-cost, i.e., ${{r}_{t}}=V{{W}_{t}}({{X}_{t}})-{{Z}_{t}}\left( D\left( {{X}_{t}} \right)-\overline{D} \right)$.

In the DDQN algorithm, the target value at step $t$, denoted by ${{y}_{t}^{T}}$, is calculated as follows
\begin{equation}\label{Q_value}
	y_{t}^{T}={{r}_{t+1}}+\gamma {{\max }_{{{a}_{t}}}}Q\left( {{s}_{t+1}},{{a}_{t}};{{\theta }^{T}} \right),
\end{equation}
where $\gamma$ is the discount factor, and ${{\theta }^{T}}$ is the network parameters of the target network. Function $Q\left( \cdot  \right)$ is calculated based on value function $U\left( \cdot  \right)$ and advantage function $A\left( \cdot  \right)$, which can be expressed as
\begin{equation}
	\begin{split}
	Q\left( {{s}_{t}},{{a}_{t}};{{\theta }_{1}},{{\theta }_{2}},{{\theta }_{3}} \right)&=U\left( {{s}_{t}};{{\theta }_{1}},{{\theta }_{2}} \right)+A\left( {{s}_{t}},{{a}_{t}};{{\theta }_{1}},{{\theta }_{3}} \right)\\&-\frac{1}{\left| \mathcal{A} \right|}\sum\nolimits_{a_{t}^{'}\in \mathcal{A}}{A\left( {{s}_{t}},a_{t}^{'};{{\theta }_{1}},{{\theta }_{3}} \right)}, 
	\end{split}
\end{equation}
where ${{\theta }_{1}}$ indicates the parameters of convolutional layers. Here, ${{\theta }_{2}}$ and ${{\theta }_{3}}$ refer to the parameters of fully-connected layers for the advantage function and the value function, respectively.  Here, $\mathcal{A}$ is the set of actions. The proposed DT-DDQN algorithm is presented in Algorithm \ref{DDQN}.

\begin{algorithm}[t]
	\caption{DT-DDQN}
	\label{DDQN}
	\textbf{Initialize} the primary network, the target network and the replay memory.
	
	\For{each episode}
	{
		Reset GoP requests, and CTDT and ETDT queues;
		
		\For{each step $t \in \{1, ..., t_{max}\}$}
		{
			
			Estimate the transcoding workload for each GoP in CTDT and ETDT queues;
			
			Update ETDT and CTDT queue lengths based on action, where ${{a}_{t}}=\arg {{\max }_{{{a}_{t}}\in \mathcal{A}}}Q\left( {{s}_{t}},{{a}_{t}};\theta  \right)$;
			
			Obtain the reward ${{r}_{t}}$ and update the state ${{s}_{t+1}}$;
			
			 Store $\left\{ {{s}_{t}},{{a}_{t}},{{r}_{t}},{{s}_{t+1}} \right\}$ into the replay memory, and sample a random mini-batch;
			
			 Calculate the target value by Eq. (\ref{Q_value});
			 
			  Calculate the loss value by $\mathcal{L}\left( \theta  \right)=\mathbb{E}\!\left[\!{{\left( {{r}_{t}}\!+\!\gamma {{\max }_{a}}Q\left( {{s}_{t+1}},{{a}_{t+1}};{{\theta }^{T}} \right)\!-\!Q\left( {{s}_{t}},{{a}_{t}};\theta  \right) \right)}^{2}} \right]$;
			  
			  Update the parameters with Adam optimizer;
			
		}
	}
\end{algorithm}

\section{Performance Evaluation}\label{Result}
In this section, simulation results are presented to evaluate the performance of the proposed DT-DDQN algorithm. We adopt a video dataset including twenty 1080p videos and eight 720p videos.\footnote{ https://media.xiph.org/video/derf/} The Matlab neural fitting toolbox with Bayesian regularization\footnote{https://www.mathworks.com/help/deeplearning/gs/fit-data-with-a-neural-network.html} is utilized to generate the BNN-based DT model for transcoding workload estimation. All input variables in Table \ref{Table1} are normalized to values with a mean of 0 and a standard deviation of 1. In addition, users’ request statistics are sampled from the public dataset \cite{data}. The main simulation parameters are presented in Table \ref{sim}.

\begin{table}[t]
	\footnotesize
	\centering
	\captionsetup{justification=centering, singlelinecheck=false}
	\caption{Simulation Parameters}
	\label{sim}
	\begin{tabular}{c c|c c}
		\hline
		\hline
		\textbf{Parameter}                 & \textbf{Value} & \textbf{Parameter}        & \textbf{Value}  \\ \hline
		Primary DQN learning rate                          & $10^{-4}$               &    $\gamma$            & 0.99 \\ 
		Target DQN learning rate & $10^{-3}$            & $\mathcal{R}$  &   5000          \\ 
		Number of episodes                        &      1000      &  $\overline{D}$   & 1.8 s          \\ 
		Number of steps       & 100 &	$\kappa$ & [1, 10] MFlops  \\
		$L_{i}^{\max }, I_{max}$ & 1.5, 0.5 s & $f$ & [5, 20] GHz \\
		 \hline
	\end{tabular}
\end{table}

We compare the proposed DT-DDQN algorithm with the
following benchmark schemes.
\begin{itemize}
	\item[$\bullet$] \textbf{\underline{R}ound \underline{R}obin (RR)}: GoPs are placed to TQs in equal portions and in a circular order.
	
	\item[$\bullet$] \textbf{\underline{P}roportional \underline{F}air (PF)}: GoPs are sequentially placed in TQs based on scheduling priority considering GoP requests and transcoding workloads.
	
	\item[$\bullet$] 
	\textbf{\underline{U}tility-based \underline{M}ulti-dimensional and \underline{M}ulti-choice \underline{K}na\underline{P}sack (UMMKP)} \cite{UMMKP}: 
	Each TQ is seen as a knapsack, while each GoP is viewed as a good. Each GoP is sequentially placed to TQs based on the user satisfaction-based utility value.
	
	\item[$\bullet$] \textbf{DDQN}: Only DDQN is adopted for the transcoding-path selection. The TWE model is a universal one without DT.
\end{itemize}

\begin{figure}[t]\label{fig4}
	\centering
	\subfigure[]{
		\label{fig4:first}
		\includegraphics[width=0.23\textwidth]{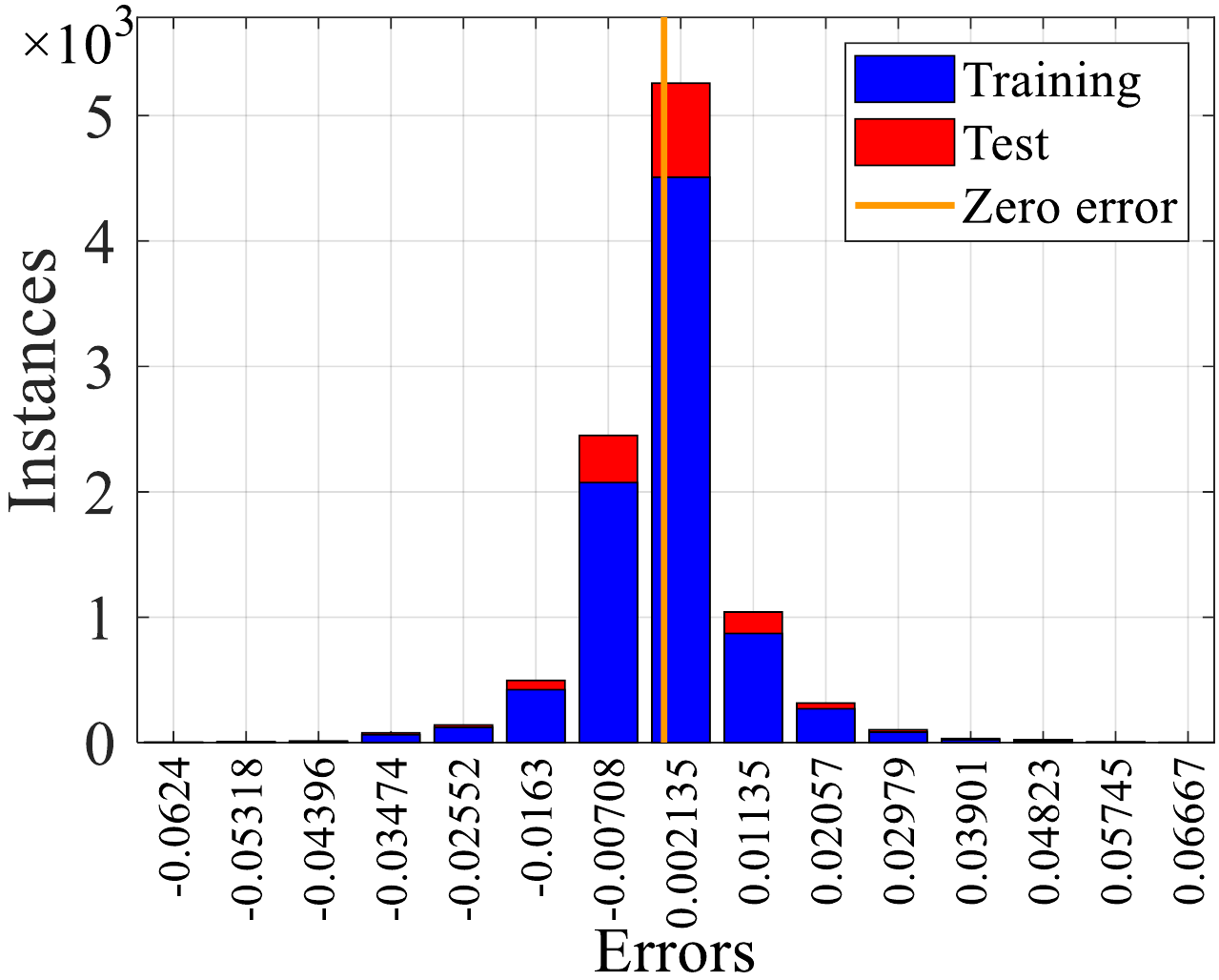}}
	\centering
	\subfigure[]{
		\label{fig4:second}
		\includegraphics[width=0.23\textwidth]{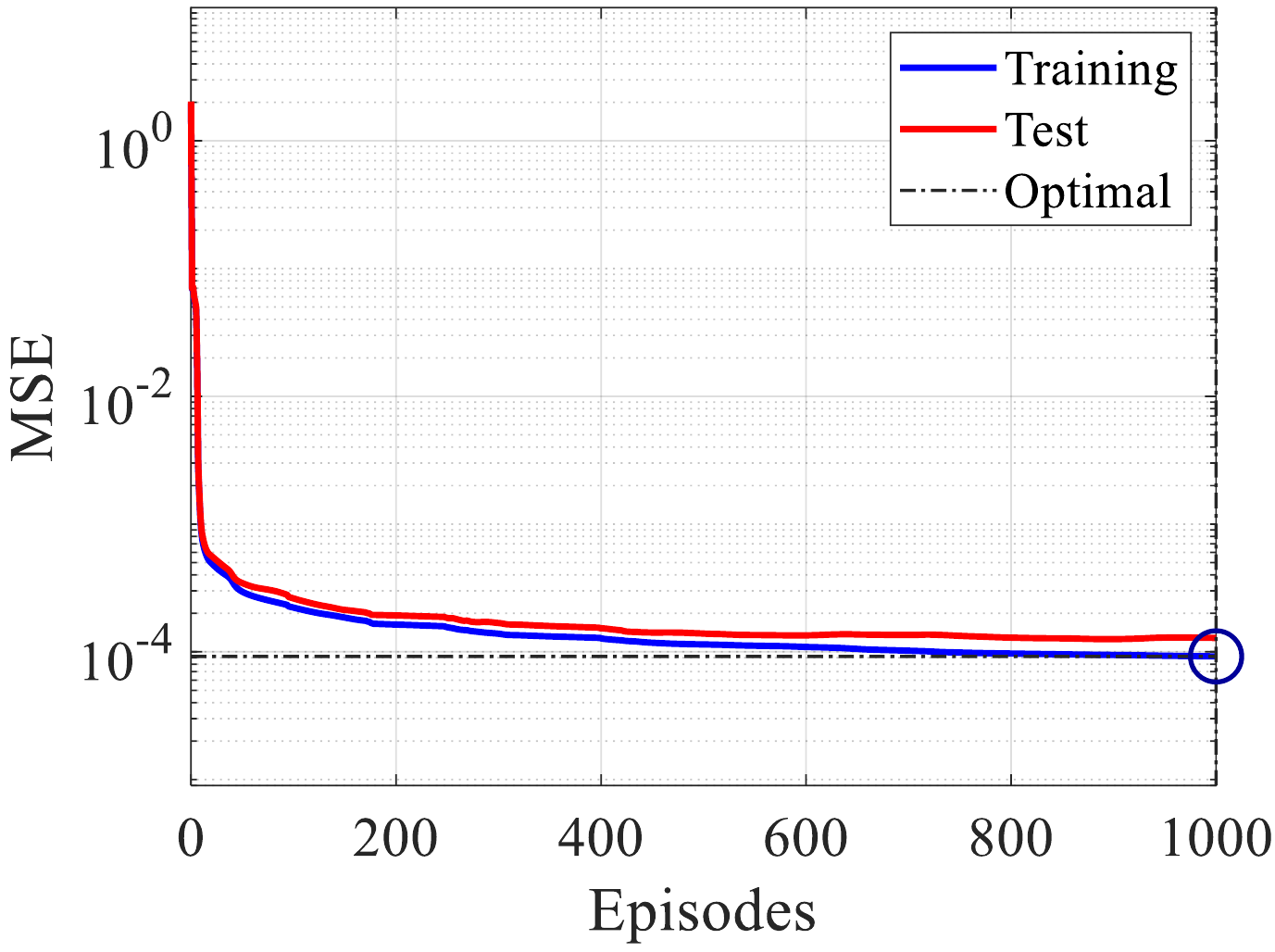}}
	\caption{BNN-based DT TWE model performance: (a) error histogram, and (b) MSE.}
	\label{fig4}
\end{figure}

In Figure \ref{fig4:first}, 7976 and 1994 instances (samples) are utilized for training and testing the BNN-based TWE models in DTs, respectively. It can be observed that the estimation errors are concentrated in [-0.0163, 0.02057], and more than half instances have fairly small errors. In Fig.~\ref{fig4:second}, the MSE can achieve a fast degradation in the initial stage and gradually converges to $10^{-4}$ when up to 1000 episodes. Therefore, the BNN-based TWE model can well extract historical transcoding data for accurate TWE. 

\begin{figure}[t]\label{fig5}
	\centering
	\subfigure[]{
		\label{fig5:first}
		\includegraphics[width=0.23\textwidth]{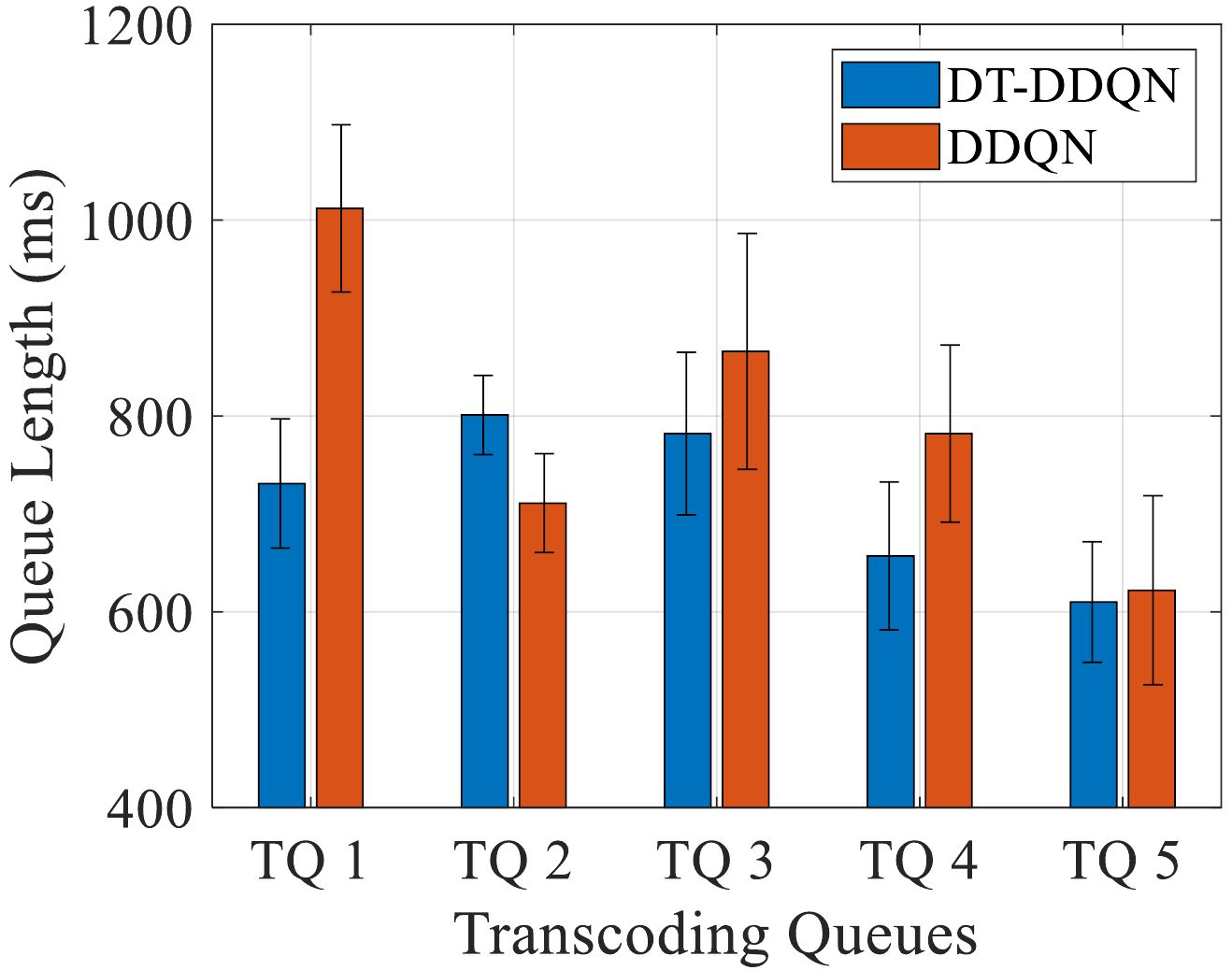}}
	\centering
	\subfigure[]{
		\label{fig5:second}
		\includegraphics[width=0.23\textwidth]{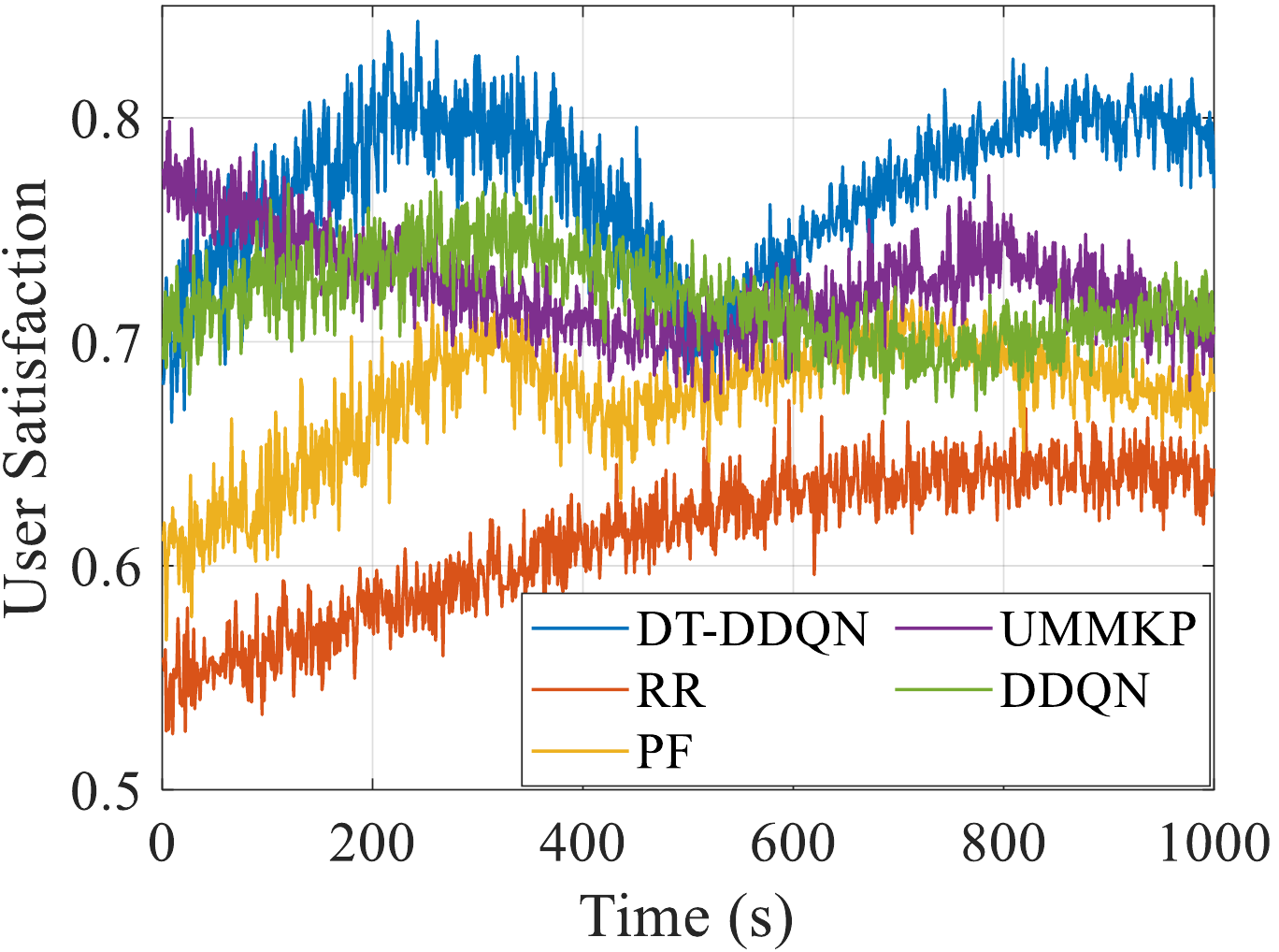}}
		\caption{Transcoding performance comparison: (a) cloud and edge TQ lengths, and (b) user satisfaction.}
	\label{fig5}
\end{figure}

Figure \ref{fig5:first} shows the average lengths and variance of five TQs in the cloud and edge servers. The TQ lengths under the proposed DT-DDQN algorithm are relatively balanced with smaller fluctuations compared with the DDQN algorithm, and the length of TQ 2 is higher. This is because the proposed DT-DDQN algorithm can accurately estimate the transcoding workload of each GoP to help the network controller balance transcoding workloads among TQs, and TQ 2 relieves a part of transcoding workloads of TQ 4. Fig.~\ref{fig5:second} shows the proposed DT-DDQN algorithm can achieve the highest user satisfaction in the most of time. Around $500\,s$, the user satisfaction of all algorithms suffers a distinct degradation since the GoP requests of high-quality version dramatically increase, and the slow TQ cannot support soaring trancoding operations of GoPs and have to transfer a part of GoPs to medium TQ and fast TQ to satisfy the service delay requirement.

\section{Conclusion}\label{Conclusion}
In this paper, we have studied a collaborative transcoding problem for better user satisfaction in live streaming. We have proposed a DT-assisted collaborative transcoding model to capture the dynamics of cloud and edge TQs and developed a DRL-based transcoding scheduling algorithm to enhance long-term user satisfaction within an average service delay threshold. The proposed DT-assisted collaborative transcoding model can also be applied to distributed computation offloading to improve the efficiency of cooperative computation. For the future work, we will investigate how to efficiently coordinate the DT-assisted cloud-edge collaborative transcoding process and wireless transmission to further enhance user satisfaction.


\bibliographystyle{IEEEtran}
\bibliography{Ref}

\begin{thebibliography}{10}
\providecommand{\url}[1]{#1}
\csname url@samestyle\endcsname
\providecommand{\newblock}{\relax}
\providecommand{\bibinfo}[2]{#2}
\providecommand{\BIBentrySTDinterwordspacing}{\spaceskip=0pt\relax}
\providecommand{\BIBentryALTinterwordstretchfactor}{4}
\providecommand{\BIBentryALTinterwordspacing}{\spaceskip=\fontdimen2\font plus
\BIBentryALTinterwordstretchfactor\fontdimen3\font minus
  \fontdimen4\font\relax}
\providecommand{\BIBforeignlanguage}[2]{{%
\expandafter\ifx\csname l@#1\endcsname\relax
\typeout{** WARNING: IEEEtran.bst: No hyphenation pattern has been}%
\typeout{** loaded for the language `#1'. Using the pattern for}%
\typeout{** the default language instead.}%
\else
\language=\csname l@#1\endcsname
\fi
#2}}
\providecommand{\BIBdecl}{\relax}
\BIBdecl

\bibitem{report}
{Grand View Research}, ``Global video streaming market share report,
  2022-2030,'' pp. 1--200, 2021.

\bibitem{Transcode}
R.~Aparicio-Pardo, K.~Pires, A.~Blanc, and G.~Simon, ``Transcoding live
  adaptive video streams at a massive scale in the cloud,'' in \emph{Proc.
  {ACM} MMSys}, 2015, Portland, OR, USA, pp. 49--60.

\bibitem{shi1}
Y.~Shi, K.~Yang, Z.~Yang, and Y.~Zhou, ``Mobile edge artificial intelligence:
  Opportunities and challenges,'' \emph{Amsterdam, The Netherlands: Elsevier},
  pp. 57--65, 2021.

\bibitem{delay}
Y.~Zheng, D.~Wu, Y.~Ke, C.~Yang, M.~Chen, and G.~Zhang, ``Online cloud
  transcoding and distribution for crowdsourced live game video streaming,''
  \emph{{IEEE} Trans. Circuits Syst. Video Technol.}, vol.~27, no.~8, pp.
  1777--1789, 2016.

\bibitem{When}
Y.~Zhu, Q.~He, J.~Liu, B.~Li, and Y.~Hu, ``When crowd meets big video data:
  Cloud-edge collaborative transcoding for personal livecast,'' \emph{{IEEE}
  Trans. Netw. Sci. Eng.}, vol.~7, no.~1, pp. 42--53, 2020.

\bibitem{shi2}
Y.~Shi, J.~Dong, and J.~Zhang, ``Low-overhead communications in {IoT}
  networks,'' \emph{Singapore: Springer}, pp. 1--10, 2020.

\bibitem{MM}
H.~Pang, C.~Zhang, F.~Wang, H.~Hu, Z.~Wang, J.~Liu, and L.~Sun, ``Optimizing
  personalized interaction experience in crowd-interactive livecast: a
  cloud-edge approach,'' in \emph{Proc. {ACM} MM}, 2018, Seoul, Korea, pp.
  1217--1225.

\bibitem{OSCAR}
A.~Erfanian, F.~Tashtarian, A.~Zabrovskiy, C.~Timmerer, and H.~Hellwagner,
  ``{OSCAR}: On optimizing resource utilization in live video streaming,''
  \emph{{IEEE} Trans. Netw. Serv. Manage.}, vol.~18, no.~1, pp. 552--569, 2021.

\bibitem{Holistic}
X.~Shen, J.~Gao, W.~Wu, M.~Li, C.~Zhou, and W.~Zhang, ``Holistic network
  virtualization and pervasive network intelligence for {6G},'' \emph{{IEEE}
  Commun. Surveys Tuts.}, vol.~24, no.~1, pp. 1--30, 2022.

\bibitem{PQoE}
X.~Huang, C.~Zhou, W.~Wu, M.~Li, and H.~Wu, ``Personalized {QoE} enhancement
  for adaptive video streaming: A digital twin-assisted scheme,'' \emph{arxiv
  preprint arXiv: 2205.04014}, 2022.

\bibitem{STI}
\BIBentryALTinterwordspacing
{ITU-T}, ``Subjective video quality assessment methods for multimedia
  applications,'' pp. 1--50, 2022. [Online]. Available:
  \url{https://www.itu.int/rec/T-REC-P.910-202207-I/en}
\BIBentrySTDinterwordspacing

\bibitem{BNN}
F.~Burden and D.~Winkler, ``Bayesian regularization of neural networks,''
  \emph{Artificial Neural Netw.}, pp. 23--42, 2018.

\bibitem{wu}
W.~Wu, C.~Zhou, M.~Li, H.~Wu, H.~Zhou, N.~Zhang, X.~Shen, and W.~Zhuang,
  ``{AI}-native network slicing for {6G} networks,'' \emph{IEEE Wireless
  Commun.}, vol.~29, no.~1, pp. 96--103, 2022.

\bibitem{Lya}
M.~J. Neely, ``Stochastic network optimization with application to
  communication and queueing systems,'' \emph{Synthesis Lectures on
  Communication Networks}, vol.~3, no.~1, pp. 1--211, 2010.

\bibitem{DDQN}
Z.~Wang, T.~Schaul, M.~Hessel, H.~Hasselt, M.~Lanctot, and N.~Freitas,
  ``Dueling network architectures for deep reinforcement learning,'' in
  \emph{Proc. ICML}, 2016, New York, NY, USA, pp. 1995--2003.

\bibitem{data}
F.~Loh, F.~Wamser, F.~Poignée, S.~Geißler, and T.~Hoßfeld, ``Youtube dataset
  on mobile streaming for internet traffic modeling and streaming analysis,''
  \emph{Scientific Data}, vol.~9, no.~1, pp. 1--12, 2022.

\bibitem{UMMKP}
J.~Liu, W.~Zhang, S.~Huang, H.~Du, and Q.~Zheng, ``{QoE}-driven {HAS} live
  video channel placement in the media cloud,'' \emph{{IEEE} Trans.
  Multimedia}, vol.~23, pp. 1530--1541, 2020.

\end{thebibliography}
\end{document}